\documentstyle[12pt]{article}
\begin {document}

 \begin {flushright}
 ULB-TH 98/01,  RI 98/01,  CERN-TH 98/05 \\
 hep-th/9801048  \\  January 1998\\
\end{flushright}
\vspace{0.5cm}
\begin {center}
 { \huge  Statistical Entropy of Schwarzschild Black Holes}\\
\vspace{1.5cm}
 {  \large  F.~Englert}\footnote{ E-mail:
fenglert@ulb.ac.be}$^{\dagger *}$
 {\large and E.~Rabinovici}\footnote{E-mail:
eliezer@vms.huji.ac.il\\}$^{\ddagger *}$

 \vspace{0.7cm} $^\dagger${\it Service de Physique Th\'eorique}\\
 {\it Universit\'e Libre de Bruxelles, Campus Plaine, C.P.225}\\
 {\it Boulevard du Triomphe, B-1050 Bruxelles, Belgium}\\
\vspace{.3cm}
 $^\ddagger${\it Racah institute of Physics }  \\
 {\it Hebrew University, Jerusalem, Israel\\
\vspace{.3cm}
$^*$Theory Division, CERN\\ CH-1211 Geneva 23, Switzerland }
 \end{center}
 \vspace{.5cm}
 \begin {abstract}

\noindent
 The entropy of a seven dimensional Schwarzschild black hole of
arbitrary large radius is obtained    by a mapping  onto  a near
extremal self-dual three-brane whose partition function can be
evaluated. The three-brane arises from duality after submitting a
neutral blackbrane,   from which the Schwarzschild black hole  can be
obtained by  compactification, to  an infinite boost  in non compact
eleven dimensional space-time   and then to a Kaluza-Klein
compactification. This limit can be   defined in precise terms and
yields the Bekenstein-Hawking value up to a factor of order one which
can be set to be exactly one with the extra assumption of keeping
only  transverse brane excitations.  The method can be generalized
to  five and four dimensional black holes.
 \end{abstract}
 \newpage

\noindent  {\bf 1. GENERAL CONSIDERATIONS}
\medskip

A major justification for indulging in the study of string theory,
brane configurations and M-theory is that it may offer a useful point
of view on quantum gravity. Indeed discoveries in D-brane physics have
enabled the  study of   problems  which lay outside the realm of
classical general relativity,  and in particular  the evaluation   of
the black hole entropy from a counting of quantum states.  Such
computations were successfully performed for extreme supersymmetric
black holes in cases where the dilaton field is finite on the horizon
\cite{M}. This is the case for charged black holes formed  from
marginally bound intersecting BPS branes in four and five space-time
dimensions \cite{SV,CM,HS,MStro,JKM}.  The correct Bekenstein-Hawking
value $A/4$ for the entropy was also obtained for near extremal black
three-branes \cite{GKP},  which also has a finite   (and  in fact
constant) dilaton field   although its area goes to zero  at
extremality.   However in  this case  the factor
$1/4$ was not obtained by direct calculation.  It was suggested
\cite{GKP} to recover it by disregarding the longitudinal excitations
of the brane.  This recipe has not been clarified; it might arise from
some peculiarity of the gauge field dynamics  or perhaps just provide
a phenomenological way to take into account the departure from
supersymmetry in a case where the entropy vanishes with the   area in
the BPS extremal limit.   Recently,  microscopic considerations based
on the Matrix model have also been applied to Schwarzschild black
holes \cite{BFKS,KS,BFKS2,HM,OZ} but hitherto   in a qualitative way.
Other considerations \cite{SS} involve a connection between the
Schwarzschild black hole and the 2+1 dimensional BTZ black hole
\cite{btz}, which has been given a microscopic description in
\cite{carlip,S,BSS}. In this note we suggest a quantitative analysis
applying some M-theory ideas.

We propose to evaluate the entropy of Schwarzschild black holes of
arbitrary large radius by mapping them through boosts and duality to
nearly extremal charged black holes.  We use the idea, first proposed
in reference \cite{DMRR}, to study properties of a Schwarzschild black
hole by viewing it  as a compactification  of a blackbrane of eleven
dimensional  supergravity and relating it to a charged black hole.
The latter is obtained  by subjecting the blackbrane to a boost in
uncompactified  space-time  followed by a Kaluza-Klein
compactification  on a different radius.  By tuning the ratio of the
two compactifications radii, it is possible to ensure that the
classical thermodynamic  entropy of the two holes are the same and we
shall argue that the equivalence extends to the counting at the
quantum level.  The extreme limit corresponds to infinite boosts and
vanishing Schwarzschild radius. We shall examine here a different
limit, which we call the near extremal limit, for which the
Schwarzschild radius remains arbitrarily large for infinite boost.
Departure in  energy from extremality are made vanishingly small
although the mass of the charged black hole obtained after
compactification of the boosted neutral blackbrane increases
indefinitely with the boost. The entropic equivalence  of the neutral
and the charged black hole remains valid for infinite boosts as a
consequence of an infinite concomitant rescaling of  the Newton
constant and of the charge quantum in the near extremal limit.

We shall test our proposal on the seven dimensional Schwarzschild
black hole,  which will be related to a  near-extremal three-brane,
and give a detailed analysis of  our limiting procedure.  Counting the
number of the three-brane excitations, one recovers the
Bekenstein-Hawking value for the entropy up to a factor of order unity
which becomes exactly one if, for example,  only the transverse
excitations of the brane are counted.  The reason of this incomplete
success is    rooted in the vanishing of the area in the extreme limit
of the three-brane which necessitates a counting of non BPS-states to
get a non vanishing entropy close to extremality. The use of
supersymmetry to relate the semi-classical entropy to a string theory
counting of states in flat space, is then, as in reference \cite{GKP},
not a priori fully justified.  This is not the case for    black holes
in four and five space-time dimensions generated by intersecting
branes which have a finite area in the extremal limit. One therefore
expects that, using the mapping procedure, complete agreement may be
reached for Schwarzschild black holes  in five and four dimensions.
However  to generate intersecting branes from Kaluza-Klein charges
requires several boosts \cite{DMRR}. The analysis of the limiting
procedure is then  more involved   and is differed to a separate
publication \cite{AEH}.

In the next section we   review the relation between the neutral
Schwarzschild black hole and the charged one and give a precise
definition of   the near extremal limit.  In section 3 this limiting
procedure is analyzed and the computation of the entropy of the seven
dimensional Schwarzschild black hole is performed.
\bigskip
\bigskip

\noindent  {\bf 2. FROM NEUTRAL TO    CHARGED BLACK HOLES. }
\medskip

The line element of a neutral black $p+1$ brane in eleven dimensional
space-time is
\begin{equation}
 -[1-({r_0\over r})^{D-3}] dt^2 +  [1-({r_0\over r})^{D-3}]^{-1} dr^2
+ r^2 d\Omega_{D-2} + dz^2 + (dx_1^2 + dx_2^2 +\ldots +  dx_p^2)
\label{blackbrane}
 \end{equation}  where $D=10-p$. When compactified on $S^1 \times
T^p$, Eq.(\ref{blackbrane}) describes a Schwarzschild black hole in $D$
space-time dimensions. Its mass $M$ and entropy $S$ are
 \begin {eqnarray}
 M&= & {1\over 16 \pi G_D} \Omega_{D-2} (D-2)    r_0^{D-3} ,
\label{mass}\\
 S&=& {1\over 4 G_D} \Omega_{D-2}   r_0^{D-2} .
 \label {entropy}
 \end{eqnarray}
 Here $r_0$ is the Schwarzschild radius and the D-dimensional Newton
constant
$G_D$ is related to the eleven dimensional Planck length $l_p$ by
\begin{equation}
  G_D = {l_p\over 2\pi  R L^p}
\label{newton}
\end{equation}
 where  $2\pi R$ is the length of the $S^1$-circle in the $z$-direction
and
$ L^p$ the volume of the $T^p$-torus.

Following reference \cite{DMRR}, let us perform a Lorentz boost with
rapidity $\alpha$ on the blackbrane in the uncompactified eleven
dimensional space-time along the $z$-direction:
 \begin {eqnarray}
 z^\prime &= & z\cosh\alpha + t\sinh\alpha\\
 t^\prime&=& t\cosh\alpha + z\sinh\alpha .
 \end{eqnarray} This boost is not a symmetry of the solution but it
will be useful to view it as a coordinate transformation. When
compactified on
$S^{1\prime} \times T^p$ where the radius of the $S^{1\prime} $ circle
in the $z^\prime$ direction is $R^\prime$, one obtains   a charged
black hole with metric
 \begin{equation}
 ds_D^2 = f^{1\over D-2}(r) \left[ -f^{-1}(r) [1-({r_0\over r})^{D-3}]
dt^{\prime 2} +  [1-({r_0\over r})^{D-3}]^{-1} dr^2 +  r^2
d\Omega_{D-2}\right]
\label{chargebh}
\end{equation}
 where
 \begin{equation}
 f(r) = 1 +  ({r_0\over r})^{D-3}\sinh^2 \alpha.
\label{boost}
\end{equation} The mass and the entropy of the charged black hole,
expressed  in terms of the new  Newton constant $G^\prime_D = l_p
/2\pi R^\prime L^p$, follow from the metric Eq.(\ref{chargebh}). One
gets
\begin{equation}
 M_{charged}= {1\over 16 \pi G^\prime_D} \Omega_{D-2} (D-3)
(\sqrt{Q^2+{1\over 4} r_0^{2(D-3)}} + {D-1\over2(D-3)}  r_0^{D-3})
\label{massch}
 \end{equation}
 where
\begin{equation}
 Q= r_0^{D-3}  \cosh\alpha \sinh\alpha .
\label{charge}
\end{equation}
 Here $Q$ is a Kaluza-Klein charge and $r_0$ plays also the role of a
parameter measuring the departure from extremality.  The entropy of
the charged black hole is
 \begin{equation}
 S_{charged}= {1\over 4G^\prime_D} \Omega_{D-2}   r_0^{D-2}\cosh
\alpha.
 \label {entropych}
 \end{equation} If one relates the   compactification radius  $R$ and
$R^\prime$ by
\begin{equation}
 R=R^\prime \cosh\alpha,
 \label{radius}
 \end{equation}  the expansion of the horizon area due to the boost is
compensated by the increase of the Newton constant
\begin{equation} G^\prime_D = G_D \cosh \alpha
\label{newtonnew}
\end {equation} and the  entropies of the neutral and charged black
hole become identical \cite{HM,DMRR}. This can be understood in the
following terms. In the non compact eleven dimensional space-time the
points to be identified by the Kaluka-Klein compactification form, in
the boosted frame, an array separated by equal space intervals $2\pi
R^\prime$ at fixed  time $t^\prime$. These points  are viewed from the
rest frame as dilated according to Eq.(\ref{radius}) but they are
separated   by    time intervals $\Delta t = 2\pi R^\prime
\sinh\alpha$.  For the neutral blackbrane at rest such time intervals
become on the horizon   instantaneous events because of the infinite
redshift.  Horizons of the two compactified solutions in ten or lower
dimensions are thus related simply by a coordinate transformation in
eleven dimensions.   If the horizons store the relevant degrees of
freedom responsible for the black hole entropy, the degrees of freedom
of the neutral and of the charged hole   should be  equivalent to one
another under the eleven dimensional diffeomorphism group.  This
suggests that the equivalence of the entropy of the neutral and
charged black hole is not merely a consequence of matching the
classical formulas Eqs.(\ref{entropy}) and  (\ref{entropych}) but
should remain valid at the level of a quantum mechanical counting of
states. We shall take as a working hypothesis that this is indeed the
case.

The inner and outer horizons of the charged black hole occur
respectively at $r=0$ and $r=r_0$. It follows from Eq.(\ref{charge})
that at fixed charge $Q$ the charged black hole becomes extremal in the
limits $r_0 \to 0$,
$\alpha \to \infty$,  with $r_0^{D-3}\sinh \alpha \cosh\alpha$ kept
fixed. We shall consider a different limit.  Labeling by $\Delta M$ the
difference in mass between   $M_{charged}$ and the extremal value
$M_{ext}$  for the same charge    one gets, to first order in
$r_0^{D-3} / Q$,
 \begin{eqnarray}
 \nonumber
 \Delta M= {1\over 16 \pi G_D^\prime} \Omega_{D-2} {1\over 2}(D-1)
r_0^{D-3}  & =&  {D-1\over  (D-3)} M_{ext}\sinh^{-1}2\alpha\\
\label{excessmass} &=& {D-1\over  2(D-2)} M\cosh^{-1}\alpha .
\end{eqnarray} We see that  $\Delta M / M_{ext} $ tends  exponentially
to zero as $  \exp (-2\alpha) $  when $\alpha \to \infty$ independently
of the value of $r_0$.    This suggests to   let $\alpha $ tend to
infinity   keeping $r_0$ finite. In this way we can approach the
extremal limit for charges $Q$ growing indefinitely and arbitrary
large Schwarzschild radii. More precisely, we shall consider the limit
as $\alpha
\to \infty$ of a sequence of charged black holes with arbitrary large
but fixed
$r_0$ obtained by first boosting  with rapidity $\alpha$ in non
compact eleven dimensions and then compactifying on the radius
$R^\prime$ defined in Eq.(\ref{radius}). This we   call the near
extremal limit. This limit may be different from the one which would
result from considering a sequence of extremal black holes with
increasing charges. Indeed, when
$\alpha$ tends to infinity at fixed $r_0$,  the entropy remains always
finite because of the compensation of the divergence in the horizon
area against the divergence of the Newton constant.  However in some
cases, as in the one considered below, the  entropy vanishes at
extremality with the area for all finite values of the charges. In
such cases, the near extremal limit is still well defined and   keeps
track of perturbative departures from extremality.
\bigskip
\bigskip

\noindent  {\bf 3.  COUNTING  STATES OF SEVEN DIMENSIONAL BLACK HOLES}
\medskip

The near extremal limit was defined in classical terms only. However
quantum theory imposes the quantization of the Kaluza-Klein charges.
When
$\Delta M$ is strictly zero, the black hole is extremal and  charge
quantization is equivalent to mass quantization. The minimal mass of a
Kaluza-Klein excitation being $1/R^\prime$, the number of charge
quanta at extremality is $N=M_{ext} R^\prime$.   As  $\Delta M$
vanishes in the near extremal limit this relation must remain true in
this limit provided N remains finite. This is indeed the case because,
as $\alpha$ goes to infinity, the  $ \exp(2\alpha) $  divergence of the
charge $Q$ in Eq.(\ref{charge}) cancels  against the product of the
divergences of Newton's constant $G^\prime_D$ in Eq.(\ref{massch}) and
of the mass quantum
$R^{\prime -1}$.  We get
\begin{equation}
 N  = {1\over 16 \pi G_D} \Omega_{D-2} (D-3) r_0^{D-3} R=  {D-3\over
D-2} M R .
\label{number}
\end{equation} For extremal charged black holes with zero horizon
area, the only way entropy could then arise in the near extremal limit
would thus be   through a finite contribution per charge quantum for
infinitesimal $\Delta M$!  Remarkably, as we shall now see, this is
exactly what M-theory predicts.

We  use the assumed existence of M-theory to reduce, under some
conditions, the computation of a Schwarzschild black hole entropy to a
counting of string  theory states in flat space.  The strategy is   to
keep the eleven dimensional Planck length $l_p$ fixed. The charged and
neutral black holes which were compactified on different eleven
dimensional radii are   described by different ten dimensional string
theories and particles carrying Kaluza-Klein charges become
D0-branes.  We use T-duality to interpret the charged black hole   as
generated from    Dp-branes.  We then extrapolate to zero curvature
and prove that only massless excitations   are present in the near
extremal limit.  Finally,  we evaluate  the D-brane degeneracy from
the effective action of the zero-mass excitations.

Such a counting of states for branes close to extremality can only be
performed when the dilaton field is not singular on the horizon
\cite{KT}. For the case of non intersecting branes considered here,
this selects  three-branes. We shall therefore consider here only the
three-brane case which is related to the seven dimensional
Schwarzschild black hole.  The computation in the limit $\alpha \to
\infty$  will reproduce the results of  reference \cite{GKP}.  But
this
 limit    requires a   detailed analysis which we now perform.

Keeping the eleven dimensional Planck length $l_p$ fixed, the relation
Eq.(\ref{radius}) means that the charged black hole is described by a
ten dimensional string theory with string length $l_s^\prime$ and
string coupling constant $g^\prime$ related to the unboosted string
parameters $l_s$ and $g$ by
 \begin{equation}
 l_s =l_s^\prime \cosh^{-1/2}\alpha ,  \qquad
g=g^\prime\cosh^{3/2}\alpha .
\label{length}
 \end{equation}  The seven dimensional Schwarzschild black hole  is
related through the boost  to the near extremal charged black hole
described by  the metric Eq.(\ref{chargebh}) with $D=7$. This metric
can also be viewed as resulting, after compactification, from charged
near extremal self-dual three-branes in ten dimensions wrapped over a
three-torus \cite{KT,DLP}. This   follows  from T-dualities. The
three-brane is formed from  $N$ coincident   D-branes, where $N$
tends, as $\alpha \to \infty$, to the value   Eq.(\ref{number}) for
$D=7$, wrapped over a   torus dual to the original one whose volume is
$ L^3$.   Massless excitations are encoded in a $U(N)$ super
Yang-Mills field theory on the dual torus.   The volume of the dual
torus
$ L_d^3$ and the string dual coupling constant $g_d$ are given by
\begin{eqnarray}
 L_d &=&{ 4\pi^2 l_s^{\prime 2}\over L}  ,  \label{size}\\
 g_d&=&g^\prime{ 8 \pi^3 l_s^{\prime 3}\over L^3} .
\label{dualcoupling}
\end{eqnarray}  Taking into account Eqs.(\ref{length}), the string
coupling $g_d $ equals  $ g 8 \pi^3 l_s^3/L^3$; it is independent of
the boost and thus stays finite when the boost parameter $\alpha$
tends to infinity. The size of the dual torus diverges as $\exp\alpha$
as seen in Eq.(\ref{size}).

Extrapolating to the weak coupling limit at arbitrary large but fixed
value of $\alpha$, we   use perturbative string theory to count states
in flat space.    As pointed out in section one, this step is not a
priori justified because we are departing from BPS states; moreover
the physics of the conformal super Yang-Mills theory in $3+1$
dimensions in the phase of unbroken scale invariance is not well
understood, except at zero gauge coupling. We shall however assume,
following common practice, that the problem is reduced to a counting
of only the transverse free massless excitations.  Our point is indeed
not to  debate about the full justification of this procedure but
rather to use it as a successful phenomenology.

The massless states can then be counted: there are
$6 N^2$ bosons and fermion  massless transverse excitations present.
 If the thermal limit were valid, the entropy would   be
\begin{equation} S_{gas}=   2^{5/ 4} 3^{-(3/4)} \pi^{1/2}  N^{1/2}
\Delta M^{3/4}  L_d^{ 3/4} .
\label{thermal}
\end{equation}
 We now see why this can lead to a finite answer in the near extremal
limit. As announced, the contribution to the entropy per charge
quantum is finite:    the vanishing of $\Delta M$ is exactly
compensated as $\alpha \to \infty$ by the divergence of $L_d$.  The
near extremal limit differs from a limit leading to the same black
hole charge  through  a sequence of extremal black holes: it  keeps
track  of brane excitations which, if  $\Delta M $  were put to zero at
the outset, would be absent. This results in a finite limiting entropy.
 \addtocounter{footnote}{-2}

The fact that the size of the dual torus increases without bound means
that the density of excitations goes to zero and hence ensures
stability against  decays which are neglected in the super Yang-Mills
description (see however \cite{malda}).  The validity of the estimate
Eq.(\ref{thermal}) requires that the inverse temperature be smaller
than the characteristic size of the dual torus, namely $T^{-1}
\ll L_d$.  The Hawking temperature $T_H$ of the charged black hole is
\begin{equation}
 T_H =(\pi r_0 \cosh \alpha)^{-1},
 \label {temp}
 \end {equation} and therefore Eq.(\ref{size})   would require $r_0
\ll l_s^2/L$, which would invalidate  the classical description  of
both the Schwarzschild black hole and the charged one in the framework
of general relativity.   To be valid, this description requires
indeed  that the  curvature at the horizon in the string frame be much
smaller than the inverse string length squared.  This condition is
simply  $r_0 \gg l_s$ for the  Schwarzschild black hole and  remains
unaltered for the charged one: in the string frame
$r_0$ gets multiplied by
$\cosh^{1/2}\alpha$ \cite{HP}, which cancels the rescaling of the
string length. The validity of the classical description is
reinstalled by taking instead of $N$ coincident branes, the
entropically favored solution consisting of branes wrapping $N$ times
over the dual torus\footnote{ this trick was used in references
\cite{MS}
\cite{BFKS} in a different but related context.}, thus breaking $U(N)$
to
$U(1)^N$ \cite{hashimoto}.   Eq.(\ref{thermal})  must now   to be
interpreted as describing $N$ (instead of $N^2$) modes on a torus of
volume  $N  L_d^3$. Using the estimate Eq.(\ref{number}), we express
the seven dimensional Newton constant in terms of string parameters
through $G_7 = G_{10}/  L^3$   with
 \begin{equation}
 G_{10} = 8\pi^6 g^2  l_s^8
\label{string}
\end{equation} and the M-theory relation $g l_s = R$. We now get $l_s^4
\ll  (L^2/R) r_0^3$ which is consistent with the classical description
and in fact justifies the use of an exact thermal distribution in the
limit of asymptotically large Schwarzschild black holes.  However a
potentially dramatic consequence of the large boost would be the onset
of massive string states, which in the limit $\alpha \to \infty$
become massless. This does not happen because, comparing the scaling
properties of the temperature Eq.(\ref{temp}) and of the string length
Eq.(\ref{length}), we see that massive string states are exponentially
suppressed in the tensionless limit $\alpha \to
\infty$ !

The computation of the gas entropy is now straightforward. Using
Eq.(\ref{string}) and   $g l_s = R$, we express  $L_d$ given by
Eq.(\ref {size}) in terms of $G_7 $:
\begin{equation}
 L_d = 2 G_7^{1/3} R^{-(2/3)} \cosh \alpha .
\label{Gsize}
\end{equation} From
Eqs.(\ref{thermal}),(\ref{excessmass}),(\ref{number}),(\ref{Gsize}),(
\ref{mass}), and the value $\pi^3$ of the five dimensional unit
sphere,   one gets, in the limit
$
\alpha \to \infty$,
\begin{equation}
 S_{gas}= {1\over 4 G_7}\Omega_{5}   r_0^{5} ,
\label{final}
 \end {equation}
 in perfect agreement with the Bekenstein Hawking value Eq.(\ref
{entropy}) for $D=7$.
\bigskip
\bigskip
\bigskip

\noindent
   {\Large \bf Acknowledgements}
 \medskip

\noindent We are very grateful to G. 't Hooft for an illuminating
discussion on the mapping procedure.  One of us (F.E.) thanks   R.
Argurio and L. Houart for many helpful conversations.  This work was
supported in part by the American-Israeli Binational Science
Foundation (BSF) and the Israeli Fund for Basic Research.

\end{document}